\documentclass[aps,showpacs,showkeys,preprint]{revtex4}
\usepackage[dvips]{graphicx}
\usepackage{latexsym}
\begin{document}

\begin{flushright}
{\tt SOGANG-MP 03/08}
\end{flushright}
\begin{flushright}
\today
\end{flushright}

\vspace*{0.05cm}

\title{Entropy of (1+1)-dimensional charged black hole \\
to all orders in the Planck length}

 \author{Yong-Wan \surname{Kim}}
   \email{ywkim65@gmail.com}
 \affiliation{Institute of Mathematical Science and
School of Computer Aided Science, \\Inje University, Gimhae 621-749,
Korea}
 \author{Young-Jai Park}
 \email{yjpark@sogang.ac.kr}
 \affiliation{Department of Physics and Mathematical Physics Group, Sogang University, Seoul 121-742, Korea}

\begin{abstract}
We study the statistical entropy of a scalar field on the
(1+1)-dimensional Maxwell-dilaton background  without an artificial
cutoff considering corrections to all orders in the Planck length
from a generalized uncertainty principle (GUP) on the quantum state
density. In contrast to the previous results of the higher
dimensional cases having adjustable parameter, we obtain an
unadjustable entropy due to the independence of the minimal length
while this entropy is proportional to the Bekenstein-Hawking
entropy.
\end{abstract}

\pacs{04.70.Dy, 04.60.Kz, 04.62.+v.}
\keywords{Generalized uncertainty relation; black hole entropy; Maxwell-dilaton}

\maketitle

\section{Introduction}
Since 't Hooft statistically obtained the Bekenstein-Hawking entropy
\cite{bek} of a scalar field outside the horizon of the
Schwarzschild black hole by introducing an artificial brick wall
cutoff in order to remove the ultraviolet divergence near the
horizon \cite{tho} and using the leading order
Wenzel-Kramers-Brillouin (WKB) approximation, this method has been
widely used to study the statistical property of bosonic and
fermionic fields in various black holes \cite{gm,wtk,kkps}. However,
there are some drawbacks in this brick wall model such as little
mass approximation, neglecting logarithmic term and taking the
infrared term $L^3$ as a contribution of the vacuum surrounding a
black hole. Solving these problems, an improved brick-wall method
(IBWM) has been introduced by taking the thin-layer outside the
event horizon of a black hole as the integral region \cite{lz}. As a
result, this IBWM has been solved these drawbacks except the
artificial cutoffs.

Recently, in Refs. \cite{li,liu,kkp}, the authors calculated the
entropy of black holes to leading order in the Planck length by
using the newly modified equation of states of density motivated by
a GUP \cite{gup,gup1}, which drastically solves the ultraviolet
divergences of the just vicinity near the horizon replacing the
brick wall cutoff with the minimal length. In particular, we have
studied the entropy of a scalar field on the (1+1)-dimensional
charged black hole background up to leading order by using the GUP
\cite{kpo}. However, Yoon et. al. have very recently pointed out
that since the minimal length $\sqrt{\lambda}$ is actually related
to the brick wall cutoff $\epsilon$, the integral about $r$ in the
range of the near horizon should be carefully treated for a
convergent entropy \cite{yoon}.

It is also well-known that the deformed Heisenberg algebra
\cite{gup} leads to a GUP showing the existence of the minimal
length \cite{kn,Moffat,spallucci,nouicer}, which originates from the
quantum fluctuation of the gravitational field \cite{ven,mag}. On
the other hand, quantum gravity phenomenology has been tackled with
effective models based on GUPs and/or modified dispersion relations
\cite{Cam,Sabine} containing the minimal length as a natural
ultraviolet cutoff \cite{fsp}. The essence of ultraviolet finiteness
of the Feynman propagator, which displays an exponential ultraviolet
cutoff of the form of $\exp \left( -\lambda p^2\right)$, can be also
captured by a nonlinear relation $p=f(k)$, where $p$ and $k$ are the
momentum and the wave vector of a particle, respectively,
generalizing the commutation relation between operators $\hat{x}$
and $\hat{p}$ to
\begin{equation}\label{xp}
[\hat{x}, \hat{p}] = i\hbar \frac{\partial p}{\partial k}~\Leftrightarrow
\Delta x \Delta p \geq \frac{\hbar}{2} \left|~ \left< \frac{\partial
p}{\partial k} \right> ~\right|
\end{equation}
at quantum mechanical level \cite{Sabine}. Recently, Nouicer has
extended the calculation of entropy to all orders in the Planck
length for the Randall-Sundrum brane case by arguing that a GUP up
to leading order correction in the Planck length is not enough
\cite{kn} because the wave vector $k$ does not satisfy the
asymptotic property in the modified dispersion relation
\cite{Sabine}. After this work, we have obtained the entropy to all
orders for the 4D Schwarzschild case by carefully considering the
integral about $r$ in the range of the near horizon \cite{kp4}. Very
recently, we have numerically calculated the entropy to all order
corrections of a scalar field on the (2+1)-dimensional DS black hole
background introducing the incomplete $\Gamma$-functions \cite{kp3}.

In this paper, we study the entropy to all order corrections in the
Planck length of a scalar field on the charged black hole background
in the lowest (1+1) dimensions carefully considering the integral
about $r$ in the range $(r_+, r_+ + \epsilon)$ near the horizon. As
expected, this study of the 2D case is non-trivial in contrast to
the 4D Schwarzschild case \cite{kp4} because we should introduce the
incomplete $\Gamma$-function and carry out numerical calculation as
the 3D de Sitter case \cite{kp3}. By using the novel equation of
states of density \cite{kn} motivated by the GUP, we calculate the
quantum entropy of a massive scalar field on the (1+1)-dimensional
charged black hole background without any artificial cutoff and
little mass approximation while satisfying the asymptotic property
of the wave vector $k$ in the modified dispersion relation. In
contrast to the previous results of the higher dimensional cases
having adjustable parameter $\alpha$~\cite{kp4,kp3}, we obtain an
unadjustable entropy while this entropy is proportional to the
Bekenstein-Hawking entropy. From now on, we take the units as
$G=\hbar=c=k_{B}\equiv 1$.

Let us now start with the two-dimensional Maxwell-dilaton action
induced by the low energy heterotic string theory \cite{mny,NO},
which is given by
\begin{equation}
\label{action} S = \frac{1}{2\pi} \int d^{2}x
\sqrt{-g}e^{-2\phi}[R+4(\bigtriangledown\phi)^2+ 4\Lambda^2-
\frac{1}{4}F^2],
\end{equation}
where $\phi$ is a dilaton field, $\Lambda^2$ a cosmological
constant, and $F$ a Maxwell field tensor. In the Schwarzschild
gauge, the metric and field tensors are assumed to be
\begin{eqnarray}
\label{metric} ds^2 &=& - f(r) dt^2 +\frac{1}{f(r)} dr^2,\\
F_{rt} &=& F_{rt}(r).
\end{eqnarray}
Simple static solutions with suitable boundary conditions are known
as follows:
\begin{eqnarray}
\label{sol} \phi(r) &=& \frac{1}{4}\ln 2 - \Lambda r,\\
F_{rt}(r) &=& \sqrt{2} Q e^{-2\Lambda r}, \\
f(r) &=& 1- \frac{M}{\Lambda} e^{-2\Lambda r}+
\frac{Q^2}{4\Lambda^2}e^{-4\Lambda r}.
\end{eqnarray}
There are two coordinate singularities $r_{\pm}$ which correspond to
the positions of the outer event horizon and the inner Cauchy
horizon:
\begin{equation}
\label{horizon} r_{\pm} = \frac{1}{2\Lambda} \ln \left[
\frac{M}{2\Lambda} \pm \sqrt{\left(\frac{M}{2\Lambda}\right)^2 -
\left(\frac{Q}{2\Lambda}\right)^2}\right],
\end{equation}
where the Cosmic censorship leads to the condition $M\geq Q$. The
Hawking temperature is given by the surface gravity \cite{ms} as
$T_{H} = \frac{\Lambda}{2\pi} [1- e^{-2\Lambda(r_{+} - r_{-})}]$.

In this Maxwell-dilaton background, let us consider a scalar field
with mass $m$ under the background (5) and (6), which satisfies the
Klein-Gordon equation $(\Box- m^{2}) \Phi = 0$. Substituting the
wave function $\Phi(r, \theta, t) = e^{-i\omega t}R(r, {\theta})$,
we find that the Klein-Gordon equation becomes
\begin{equation}
\label{rtheta0} \frac{d^{2} R}{{dr}^2}  + \frac{1}{f} \frac{df}{dr}
\frac{dR}{dr} + \frac{1}{f} \left(\frac{\omega^{2}}{f} - m^{2}
\right)R = 0.
\end{equation}

By using the leading order WKB approximation \cite{tho}
with $R \sim exp[iS(r,\theta)]$, we have
\begin{equation}
\label{wkb} {p_{r}}^{2} = \frac{1}{f}\left(\frac{\omega^{2}}{f} -
m^{2}\right),
\end{equation}
where $p_{r} = dS/dr$. On the other hand, we also have the square
module momentum
\begin{equation}
\label{smom} p^{2} = p_{r}p^{r} = g^{rr}{p_{r}}^{2} = f{p_{r}}^{2} =
\frac{\omega^{2}}{f} - m^{2}
\end{equation}
with the condition $\omega\geq m \sqrt{f}$.

Considering the modified dispersion relation (\ref{xp}), the usual
momentum measure $dp_r$ is deformed to be $dp_r \frac{\partial
k_r}{\partial p_r}$. Then, according to the Refs.
\cite{spallucci,nouicer}, we have
\begin{equation}
\frac{\partial p_r}{\partial k_r}=~e^{\lambda p^{2}}, \label{measure}
\end{equation}%
where $\lambda$ is a constant of order one in the Planck length
units having the dimension of 1/momentum squared. Note that in the
limit of $\lambda \rightarrow 0$, we recover the usual Heisenberg
commutation relation. Moreover, this particular type of the
nonlinear dispersion relation is invertible and satisfies the
requirement that for small energies than the cutoff the usual
dispersion relation is recovered, while for large energies the wave
vector asymptotically reaches the cutoff. These
criteria~\cite{Sabine} for selecting a proper Ansatz among different
alternatives~\cite{var} have been also investigated in various
context in phenomenological consequences~\cite{gup}. Then, the
deformed algebra, which is given by $\left[ X,P\right] =i~e^{\lambda
P^{2}}$ with the representations $X \equiv i~e^{\lambda P^{2}}
{\partial _{p}}$ and $P \equiv p$ of the position and momentum
operators, respectively, leads to the generalized uncertainty
relation including all order corrections as
\begin{equation}
\Delta X \Delta P \geq \frac{1}{2}\left\langle e^{\lambda P^{2}}
\right\rangle \geq \frac{1}{2} e^{\lambda \left( \left( \Delta
P\right)^{2}+\left\langle P\right\rangle ^{2}\right)}. \label{xp1}
\end{equation}
Note that $\left\langle P^{2n}\right\rangle \geq \left\langle
P^{2}\right\rangle ^{n}$ and $\left( \Delta
P\right)^{2}=\left\langle P^{2}\right\rangle -\left\langle
P\right\rangle^{2}$.

Next, in order to investigate the quantum implications of this
deformed algebra, let us solve the above relation (\ref{xp1}) for
$\Delta P$ that is satisfied with the equality sign. Then, the
momentum uncertainty is simply given by
\begin{equation}
\Delta P =\frac{e^{\lambda \langle P \rangle^2 }}{2\Delta X}
e^{\lambda \left(\Delta P\right)^{2}}. \label{argu}
\end{equation}
On the other hand, if we define $W(\xi)\equiv -2\lambda \left(\Delta
P\right)^{2}$ with $\xi=\frac{\lambda}{2(\Delta X)^2} e^{-2\lambda
\left\langle P\right\rangle ^{2}}$, we obtain the relation $W\left(
\xi\right) e^{W\left( \xi\right) }=\xi$, which is just the
definition of the Lambert function \cite{Lambert}.
In order to have a real physical solution for $\Delta P$,
the argument of the Lambert function is required
to satisfy $\xi \geq -1/e$, which naturally leads to the position
uncertainty as $\Delta X \geq \sqrt{{e\lambda}/2}~e^{\lambda \langle P
\rangle^2} \equiv \Delta X _{\min}$.
Here, $\Delta X_{\min }$ is a minimal uncertainty in position.
Moreover, this minimal length intrinsically derived for physical
states with $\left\langle P\right\rangle =0$ is given by
$\Delta X^A_{0}=\sqrt{{e\lambda}/2}$,
which is the absolutely smallest uncertainty in position. In fact,
this minimal length plays a role of the brick wall cutoff
effectively giving the thickness of the thin-layer near the
horizon \cite{li,liu,kkp,kpo}.
Note that the minimal length up to the leading order in a series expansion of
Eq. (\ref{argu}) around $\left\langle P\right\rangle=0$ is given by $\Delta
X^L_0=\sqrt{\lambda}~<~\Delta X^A_{0}$, where the superscripts
denote the leading order ($L$) and all orders ($A$), respectively.
However, only this leading order correction of the GUP does not
satisfy the property that the wave vector $k$ asymptotically reaches
the cutoff in large energy region as recently reported in Ref.
\cite{Sabine}.

Now, let us calculate the statistical entropy of a scalar field on
the (1+1)-dimensional charged black hole background to all orders in
the Planck length. When gravity is turned on, the number of quantum
states in a volume element in phase cell space based on the GUP in
the (1+1) dimensions is given by $dn_A = \frac{dr
dp_r}{2\pi}e^{-\lambda p^2}$, where $p^2 = p^{r}p_{r}$ and one
quantum state corresponding to a cell of volume is changed from
$2\pi $ into $2\pi e^{\lambda p^2}$ in the phase space
\cite{li,liu}. Then, the number of quantum states with energy less
than $\omega$ is given by
\begin{eqnarray}
\label{Allnqs} n_{A}(\omega) &=& \frac{1}{2\pi} \int dr
dp_{r} e^{- \lambda p^2} \nonumber   \\
&=& \frac{1}{\pi}\int dr \frac{1}{\sqrt{f}}
\left(\frac{{\omega}^2}{f}- m^{2}\right)^{\frac{1}{2}} e^{- \lambda
(\frac{{\omega}^2}{f}- m^{2})}.
\end{eqnarray}

 On the other hand, for the bosonic case the free energy at inverse
temperature $\beta$ is given by
\begin{equation}
\label{def} F_A = \frac{1}{\beta}\sum_K \ln \left[ 1 - e^{-\beta
\omega_K} \right],
\end{equation}
where $K$ represents the set of quantum numbers. By using Eq.
(\ref{Allnqs}), the free energy can be rewritten as
\begin{eqnarray}
\label{LfreeE0}
 F_A~& \approx &~\frac{1}{\beta} \int dn_{A}(\omega) ~\ln
            \left[ 1 - e^{-\beta \omega} \right]  \nonumber   \\
   &=& - \int^{\infty}_{\mu\sqrt{f}} d \omega \frac{n_A(\omega)}{e^{\beta\omega} -1} \nonumber   \\
  &=& - \frac{1}{\pi} \int^{r_{+}+\epsilon}_{r_{+}} dr
\frac{1}{f} \int^{\infty}_{0} d\omega \frac{\omega}{(e^{\beta
\omega} -1)}
 e^{- \lambda {\frac{{\omega}^2}{f}}}.
\end{eqnarray}
Here, we have taken the continuum limit in the first line and
integrated by parts in the second line. Furthermore, in the last
line of Eq. (\ref{LfreeE0}), since $f \rightarrow 0$ near the event
horizon, {\it i.e.}, in the range of $(r_+, r_+ + \epsilon)$,
${{\omega}^2}/f
   - \mu^{2}$ becomes ${\omega}^2/f$
although we do not require the little mass approximation.

Moreover, we are only interested in the contribution from the just
vicinity near the horizon, $(r_+, r_+ +\epsilon)$, which corresponds
to a proper distance of order in the minimal length,
$\sqrt{{e\lambda}/2}$. This is because the entropy closes to the
upper bound only in this vicinity, which it is just the vicinity
neglected by the brick wall method \cite{tho,gm,kkps}. Then, we have
\begin{eqnarray}
\label{invariant} \sqrt{\frac{e\lambda}{2}}=\int^{r_+
+\epsilon}_{r_+} \frac{dr}{\sqrt{f(r)}}
                \approx \sqrt{\frac{2\epsilon}{\kappa}},
\end{eqnarray}
where $\kappa$ is the surface gravity at the horizon of the black
hole and it is identified as
$\kappa=\frac{1}{2}\frac{df}{dr}|_{\beta =\beta_{H}} = 2\pi
{\beta_H}^{-1}$. Note that the Taylor's expansion of $f(r)$ near the
horizon is given by $f(r) \approx  2 \kappa (r-r_{+})+ {\cal
O}\left( (r-r_{+})^2 \right)$.

Before calculating the entropy, let us mention that Yoon et. al.
have recently suggested that since the minimal length
$\sqrt{\lambda}$ in Eq. (\ref{invariant}) is related to the brick
wall cutoff $\epsilon$, the integral about $r$ in the range of the
near horizon should be carefully treated for a convergent entropy
\cite{yoon}. In particular, although the term $(e^{\beta \omega}-1)$
in Eq. (\ref{LfreeE0}) with $x= \sqrt{\frac{\lambda}{f}}\omega$ was
expanded in the previous works giving
$\beta\sqrt{\frac{f}{\lambda}}x$, one may not simply expand up to
the first order because since $0\leq \frac{f}{\lambda} =
\frac{2\kappa(r-r_{+})}{\lambda} \leq \frac{2 \kappa
\epsilon}{\lambda} = \kappa^2$ near the horizon~\cite{kp4}.

Now, let us carefully consider the integral about $r$ near the
horizon by extracting out the $\epsilon$-factor through the Taylor's
expansion of $f(r)$. Then, the free energy of $F_A$ in Eq.
(\ref{LfreeE0}) can be written as
\begin{equation}
\label{LfreeEnoMass}
  F_A \approx - \frac{1}{\pi}
   \int^{\infty}_{0} d\omega
   \frac{\omega}{(e^{\beta \omega} -1)}
  \Lambda_A(\omega,\epsilon),
\end{equation}
where $\Lambda_A$ is defined by
\begin{equation}
\label{Lam}
  \Lambda_A \equiv
 \int^{r_{+} +\epsilon}_{r_{+}} dr \frac{1}{2\kappa(r-r_+)}
   e^{- \frac{\lambda {\omega}^2}{2\kappa(r-r_{+})}}.
\end{equation}
By defining $t=\frac{\lambda {\omega}^2}{2\kappa(r-r_{+})}$,
$\Lambda_A(\omega,\epsilon)$ becomes
\begin{equation}
\label{LfreeEnoMass1}
 \Lambda_A =  \frac{1}{2\kappa}
    \int^{\infty}_{\xi} dt \frac{1}{t} e^{- t} =  \frac{1}{2\kappa} \Gamma(0,~\xi),
\end{equation}
where we have used the incomplete $\Gamma$-function given as
$\Gamma(a,\xi)=\int^{\infty}_{\xi}t^{a-1} e^{-t} dt$ with $\xi
\equiv \frac{\lambda {\omega}^2}{2\kappa
   \epsilon}$.
Then, the all order corrected entropy through the GUP is given by
\begin{equation}
\label{finalS}
  S_A = \beta^2 \frac{\partial F_A}{\partial\beta}\mid_{\beta=\beta_H}  \approx  \left(\frac{16}{\pi^2} \delta \right)
  \frac{1}{2} > \frac{1}{4}(A_2) = \frac{1}{2},
\end{equation}
where the numerical value of $\delta$ with $y\equiv \beta\omega/2$
is given by
\begin{equation}
\label{del}
  \delta = \int^{\infty}_{0} dy \frac{y^2}{\sinh^{2}{y}} \Gamma\left(0,~\frac{2y^2}{e \pi^2}\right)\approx
  4.66,
\end{equation}
and $A_2$ denotes the 2D area. This is the all order corrected
finite entropy based on the GUP. We note that all order GUP
corrected entropy does not give a logarithmic correction in the
leading order WKB approximation~\cite{Log}.

Now, it seems appropriate to comment on the entropy (\ref{finalS}),
which is obtained through the Taylor expansion of $f(r)$ to all
orders in the Planck length. Since $A_2 = 2$ from the relation of
the area $A_d$ of a sphere of radius $r$ in $d$ spatial dimensions
as $ A_d = 2 \pi^{d/2} r^{d-1}/\Gamma(\frac{d}{2})$, we have failed
to obtain the exact Bekenstein-Hawking entropy $S=\frac{1}{2}$ in
contrast to the higher dimensional cases. The reason is that there
is no adjustable parameter, which is actually the minimal length
$\lambda$, in the two dimensions in contrast to the higher
dimensional cases in which all order correction faction
$\sqrt{\frac{e}{2}}$ contained in the enlarged minimal length can be
absorbed by adjusting the parameter $\alpha$ as in
Ref.\cite{kp4,kp3}.

In summary, by using the generalized uncertainty principle, we have
investigated the entropy to all orders in the Planck length of the
massive scalar field within the just vicinity near the horizon of a
static black hole in the (1+1)-dimensional charged black hole
background by carefully considering the integral about $r$ in the
range $(r_+, r_+ + \epsilon)$ near the horizon without any
artificial cutoff and little mass approximation and satisfying the
asymptotic property of the wave vector $k$ in the modified
dispersion relation. In contrast to the previous results of the
higher dimensional cases having adjustable parameter, we have
obtained an unadjustable entropy due to the independence of the
minimal length while the entropy is proportional to the
Bekenstein-Hawking entropy.

\begin{acknowledgments}
Y.-W. Kim was supported by the
Korea Research Foundation Grant funded by Korea Government
(MOEHRD): KRF-2007-359-C00007. Y.-J. Park was supported by
the Korea Science and Engineering Foundation (KOSEF) grant
funded by the Korea government (MOST) (R01-2007-000-20062-0).
\end{acknowledgments}

\end{document}